\documentclass[final,1p,times]{elsarticle}

\usepackage{amssymb}

\journal{Nuclear Physics A}

\begin{document}

\begin{frontmatter}



\title{Five-quark components in baryons}


\author{Bing-Song Zou}

\address{Institute of High Energy Physics, CAS, P.O.Box 918, Beijing
100049, China ; \\
Theoretical Physics Center for Science Facilities, CAS, Beijing
100049, China;\\
Center of Theoretical Nuclear Physics, National Lab of Heavy Ion
Accelerator, Lanzhou 730000, China}

\begin{abstract}
Evidence has been accumulating for the existence of significant
intrinsic non-perturbative five-quark components in various baryons.
The inclusion of the five-quark components gives a natural
explanation of the excess of $\bar d$ over $\bar u$, significant
quark orbital angular momentum in the proton, the problematic mass
and decay pattern of the lowest $1/2^-$ baryon nonet, etc.. A
breathing mode of $qqq\leftrightarrow qqqq\bar q$ is suggested for
the lowest $1/2^-$ baryon octet. Evidence of a predicted member of
the new scheme, $\Sigma^*(1/2^-)$ around 1380 MeV, is introduced.
\end{abstract}


\end{frontmatter}


\section{5-quark components in the proton} \label{Sect1}

In the classical quark models prior to QCD, each proton was regarded
as composed of two $u$ quarks and one $d$ quark. The classical quark
models gave very good description of the mass pattern and magnetic
moments for the baryon SU(3) baryon $1/2^+$ octet and $3/2^+$
decuplet of spatial ground states. Then the QCD introduces the gluon
field which can fluctuate into sea quark $q\bar q$ pairs in addition
to the three valence quarks inside a proton. But until about the
year of 1992, $\bar u(x)=\bar d(x)$ and $s(x)=\bar s(x)$ were
assumed for the sea quark parton distribution functions.

However, evidence has been accumulating for the existence of
significant intrinsic non-perturbative five-quark components in the
proton. A surprisingly large asymmetry between the $\bar u$ and
$\bar d$ sea quark distributions in the proton with $\bar d-\bar
u\sim 0.12$ was observed from deep inelastic scattering and
Drell-Yan experiments~\cite{Garvey}. A popular explanation for the
excess of $\bar d$ over $\bar u$ in the proton was given by
meson-cloud model~\cite{Thomas} by including in the proton a mixture
of $n\pi^+$ with the $\pi^+$ composed of $u\bar d$. Recently, by
studying strangeness in the proton, a new configuration of
diquark-diquark-antiquark structure is proposed for the five-quark
components in the proton~\cite{zou1}. The diquark cluster
configurations also give a natural explanation for the excess of
$\bar d$ over $\bar u$ in the proton with a mixture of $[ud][ud]\bar
d$ component in the proton~\cite{zou1}. The measured value of $\bar
d-\bar u\sim 0.12$ demands the 5-quark components in the proton to
be more than 12\%.

Further evidences came from proton spin problem and single-spin
asymmetry problem, which both indicate the presence of quark orbital
angular momentum in the proton. The five-quark $uudq\bar{q}$
components in the proton with the $\bar q$ of negative intrinsic
parity demand either a quark or anti-quark in the P-wave to make up
the proton of positive intrinsic parity. So it gives naturally the
nonzero quark orbital angular momentum and hence a natural
explanation to both problems~\cite{zou1,weifx}.

To tell the relative importance of meson cloud and diquark cluster
mixtures in the proton, more precise experiments on the strangeness
in the proton are needed. While the $K\Lambda$ meson cloud picture
predicts strangeness magnetic moment and strangeness radius both
negative~\cite{beck2001}, the diquark cluster picture predicts both
positive~\cite{zou1}. Present experiments are not precise enough to
distinguish the two pictures~\cite{young,baunack}. The extremely
smallness of these strangeness observables may either indicate very
small percentage of the strangeness component in the proton or about
equally the $K\Lambda$ meson cloud and $[ud][us]\bar s$
contributions canceling each other.

\section{New scheme for $N^*(1535)$ and its $1/2^-$ nonet partners with large 5-quark components} \label{Sect2}

In the simple 3q constituent quark model, the lowest spatial excited
baryon is expected to be a ($uud$) $N^*$ state with one quark in
orbital angular momentum $L=1$ state, and hence should have negative
parity. Experimentally, the lowest negative parity $N^*$ resonance
is found to be $N^*(1535)$, which is heavier than two other spatial
excited baryons: $\Lambda^*(1405)$ and $N^*(1440)$. This is the
long-standing mass reverse problem for the lowest spatial excited
baryons.

Recently a large value of $g_{N^*(1535)K\Lambda}$ is
deduced~\cite{liubc,gengls} by a simultaneous fit to BES data on
$J/\psi\to\bar pp\eta$, $pK^-\bar\Lambda+c.c.$, and COSY data on
$pp\to pK^+\Lambda$. There is also evidence for large
$g_{N^*(1535)N\eta^\prime}$ coupling from $\gamma p \to
p\eta^\prime$ reaction at CLAS~\cite{etap} and $pp\to pp\eta^\prime$
reaction~\cite{caox}, and large $g_{N^*(1535)N\phi}$ coupling from
$\pi^- p \to n\phi$, $pp\to pp\phi$ and $pn\to d\phi$
reactions~\cite{xiejj1,caox2009}, but smaller coupling of
$g_{N^*(1535)K\Sigma}$ from comparison of $pp\to p K^+\Lambda$ to
$pp \to p K^+\Sigma^0$~\cite{Sibir}.

The mass reverse problem can be easily understood by considering
5-quark components in them~\cite{liubc,zhusl}. The $N^*(1535)$ could
be the lowest $L=1$ orbital excited $|uud>$ state with a large
admixture of $|[ud][us]\bar s>$ pentaquark component having $[ud]$,
$[us]$ and $\bar s$ in the ground state.  The $N^*(1440)$ could be
the lowest radial excited $|uud>$ state with a large admixture of
$|[ud][ud]\bar d>$ pentaquark component having two $[ud]$ diquarks
in the relative P-wave. While the lowest $L=1$ orbital excited
$|uud>$ state should have a mass lower than the lowest radial
excited $|uud>$ state, the $|[ud][us]\bar s>$ pentaquark component
has a higher mass than $|[ud][ud]\bar d>$ pentaquark component. The
lighter $\Lambda^*(1405)1/2^-$ is also understandable in this
picture. Its main 5-quark configuration is $|[ud][us]\bar u>$ which
is lighter than the corresponding 5-quark configuration
$|[ud][us]\bar s>$ in the $N^*(1535)1/2^-$.

The large mixture of the $|[ud][us]\bar s>$ pentaquark component in
the $N^*(1535)$ may also explain naturally its large couplings to
the $N\eta$, $N\eta^\prime$, $N\phi$ and $K\Lambda$ together with
its small couplings to the $N\pi$ and $K\Sigma$.  In the decay of
the $|[ud][us]\bar s>$ pentaquark component, the $[ud]$ diquark with
isospin $I=0$ is stable and keeps unchanged while the $[us]$ diquark
is broken to combine with the $\bar s$ to form either $K^+(u\bar
s)\Lambda([ud]s)$ or $\phi(s\bar s)p([ud]u)$.

\begin{figure}[htb]
\vspace{-1cm}
\includegraphics[scale=0.7]{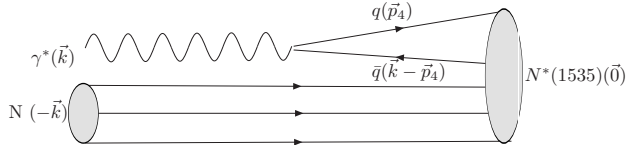}
\vspace{-17.5cm}\caption{$\gamma^{*}\to q\bar{q}$ mechanism for the
$\gamma^*N\to N^*(1535)$ transition.} \label{fig1}
\end{figure}

The inclusion of the large 5-quark components in the $N^*(1535)$
causes a natural cancelation between the contributions of the $qqq$
and $qqqq\bar q$ components to the axial charge of the N(1535)
resonance~\cite{an-riska} and introduces an important new mechanism
as shown in Fig.~\ref{fig1} for its electromagnetic transition
$\gamma^{*}N\to N^{*}(1535)$~\cite{an2009}.

\begin{figure}[htb]
\begin{minipage}[t]{70mm}
 {\includegraphics*[scale=0.6]{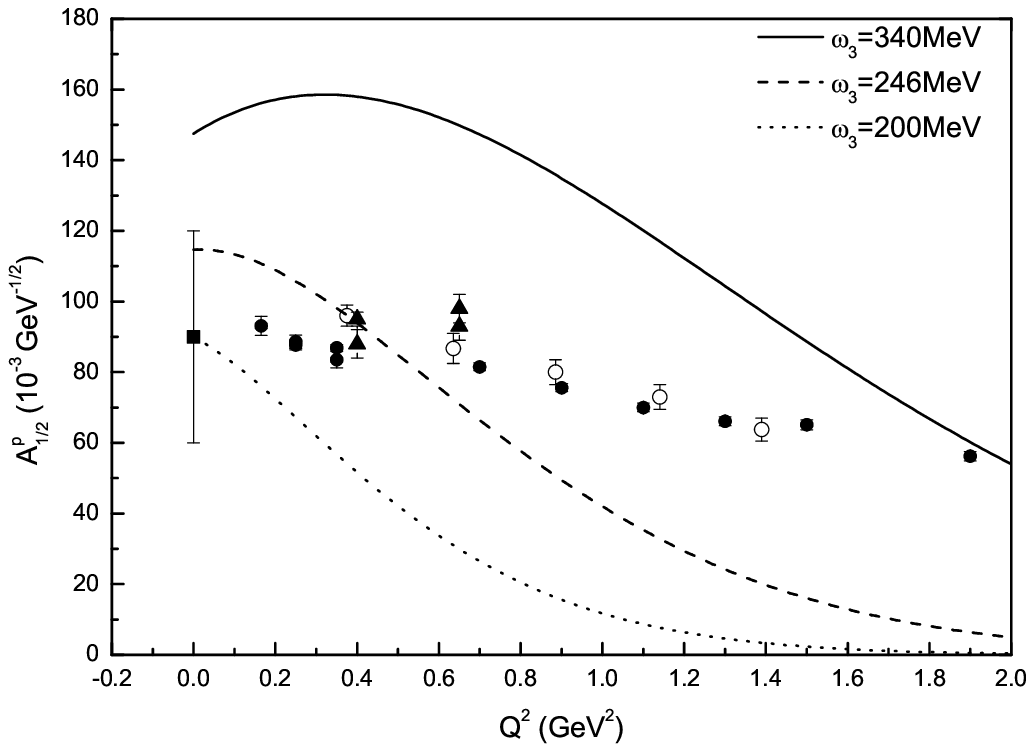}}\vskip -0cm
\end{minipage}
\begin{minipage}[t]{70mm}
{\includegraphics*[scale=0.6]{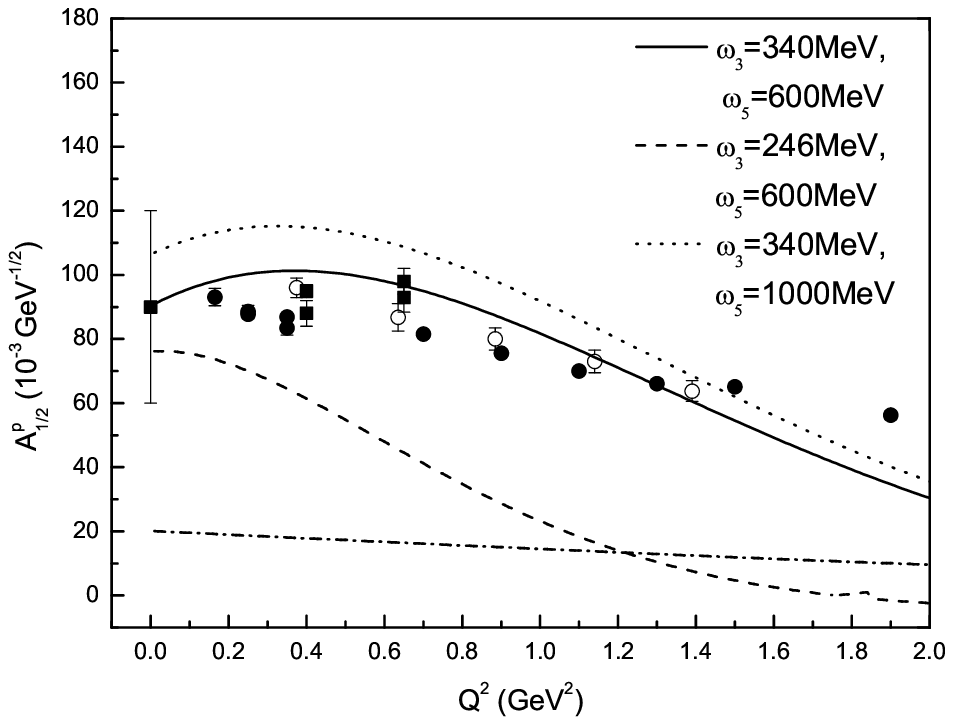}}\vskip -0cm
\end{minipage}
\caption{The helicity amplitude $A^{p}_{1/2}$ for $\gamma^{*}p\to
N^{*}(1535)$ with pure $qqq$ configuration (left) and with 5-quark
component in addition (right) where the dot-dashed curve is the
contribution from the $\gamma^{*}\to q\bar{q}$
transition~\cite{an2009}.} \label{fig:N1535}
\end{figure}

The experimental measured $\gamma^{*}N\to N^{*}(1535)$ amplitude has
a much flatter $Q^2$-dependence than the classical $qqq$ quark model
predictions as shown in Fig.~\ref{fig:N1535}(left). The new
mechanism gives a much flatter $Q^2$-dependence as shown by the
dot-dashed curve in Fig.~\ref{fig:N1535}. With about 45\% 5-quark
components in $N^*(1535)$ and 20\% in proton, the measured
electromagnetic transition $\gamma^{*}N\to N^{*}(1535)$ amplitude is
much better fitted as shown in Fig.~\ref{fig:N1535}(right).

Taking the $qqqq\bar{q}$ components into account, the constituent
quarks masses should be a bit smaller than the ones employed in the
case of pure $qqq$ configuration. To reproduce the mass for the
nucleon when the five-quark components have been included, the
values $m_{u}=m_{d}=m=290$ MeV and $m_{s}=430$ MeV are used. The
oscillator parameter for the $qqq$ components, $\omega_{3}$, is from
literature, basically determined from the proton radius.  The
oscillator parameter for the $qqqq\bar q$ components, $\omega_{5}$,
is treated as a free parameter, which is found to be larger than
$\omega_{3}$. This is consistent with other empirical evidence
favoring larger value of the
$\omega_{5}$~\cite{riska,riska1,riska2,me}. In our extended quark
model with each baryon as a mixture of the three-quark and
five-quark components, the two components represent two different
states of the baryon. For the $qqqq\bar{q}$ state, there are more
color sources than the $qqq$ state, and may make the effective
phenomenological confinement potential stronger. For the lowest
$1/2^-$ baryon octet, the larger $\omega_{5}$ value corresponds a
smaller size for the 5-quark components. An intuitive picture for
our extended quark model is the $qqq\leftrightarrow qqqq\bar q$
breathing mode like this: the $qqq$ state with $L=1$ and higher
kinetic energy has weaker potential; when quarks expand, a
$q\bar{q}$ pair is pulled out and results in a $qqqq\bar{q}$ state
with $L=0$ and stronger potential; the stronger potential leads
$qqqq\bar{q}$ state shrinking to a more compact state which then
makes the $\bar{q}$ to annihilate with a quark easily and transits
to the $qqq$ state with $L=1$ and more kinetic energy to expand;
this leads to constantly transitions between these two states.

If this picture of large 5-quark mixture is correct, there should
also exist the SU(3) nonet partners of the $N^*(1535)$ and
$\Lambda^*(1405)$, {\sl i.e.}, an additional $\Lambda^*~1/2^-$
around 1570 MeV, a triplet $\Sigma^*~1/2^-$ around 1360 MeV and a
doublet $\Xi^*~1/2^-$ around 1520 MeV~\cite{zhusl}. There is no hint
for these baryon resonances in the PDG tables~\cite{pdg}. However,
as pointed out in Ref.~\cite{zou07}, there is in fact evidence for
all of them in the data of $J/\psi$ decays and it can be easily
checked with the high statistics BESIII data in near
future~\cite{BES3-yb}. Recently, new evidence for the predicted
$\Sigma^{*}$ resonance with $J^P=1/2^-$ and mass around 1380 MeV was
found in the old data of $K^-p\to\Lambda\pi^+\pi^-$
reaction~\cite{wujj}, as introduced in the next section.

\section{Evidence for the predicted $\Sigma^*(1/2^-)$ of the new scheme} \label{Sect3}

The unquenched models give interesting predictions for the isovector
partner of the $\Lambda^*(1405)$ and $N^*(1535)$. While the
penta-quark models~\cite{zhusl,Helminen} predict a $\Sigma^*(1/2^-)$
resonance with a mass around or less than its corresponding
$\Lambda^*$ partner, the meson cloud model~\cite{jido} predicts it
to be non-resonant broad structure. The predictions of these
unquenched models and the classical quenched quark models are
distinctive and need to be checked by experiments. Hence we
re-examined the old data of $K^-p\to\Lambda\pi^+\pi^-$ reaction to
see whether there is evidence for its existence or not~\cite{wujj}.

The $K^-p\to\Lambda\pi^+\pi^-$ reaction was studied extensively
around 30 years ago for extracting properties of the
$\Sigma^*(1385)$ resonance, with $K^-$ beam momentum ranging from
0.95 GeV to 8.25 GeV. In the invariant mass spectrum of $\Lambda
\pi$ of this reaction there is a strong peak with mass around 1385
MeV and width around 40 MeV. The mass fits in the pattern of SU(3)
baryon decuplet of $J^P=3/2^+$ predicted by the classical quark
model perfectly. The angular distribution analyses also conclude
that the spin of this resonance is 3/2. However we found that all
these analyses are in fact assuming that there is only one resonance
under the peak. Nobody has considered that there are probably two
resonances there. This may be because there are no other predicted
$\Sigma^*$ resonances around this mass region in the classical quark
models. Since now a new $\Sigma^{*}$ with the $J^P=1/2^-$ around
this mass region is predicted by various unquenched models, the old
$K^-p\to\Lambda\pi^+\pi^-$ reaction should be re-scrutinized
carefully.

We examined previous experimental analyses on the
$K^-p\to\Lambda\pi^+\pi^-$ reaction. Among them, we find that the
invariant mass spectra of $\Lambda \pi^{-}$ with beam momentum
$P_{K^-}=1.0\sim 1.8 GeV$ are different from others. The peak around
$\Sigma^{*}(1385)$ in these mass spectra cannot be fit as perfect as
other sets of data with a single Breit-Wigner resonance. Since
Ref.\cite{pr69} presents the largest data sample and the most
transparent angular distribution analysis of this reaction, we
re-fit the $\Lambda\pi^-$ mass spectrum and angular distribution of
Ref.\cite{pr69} by taking into account the possibility of two
$\Sigma^*$ resonances in this mass region.

\begin{figure}[htbp] \vspace{-0.5cm}
\begin{center}
\includegraphics[width=0.49\columnwidth]{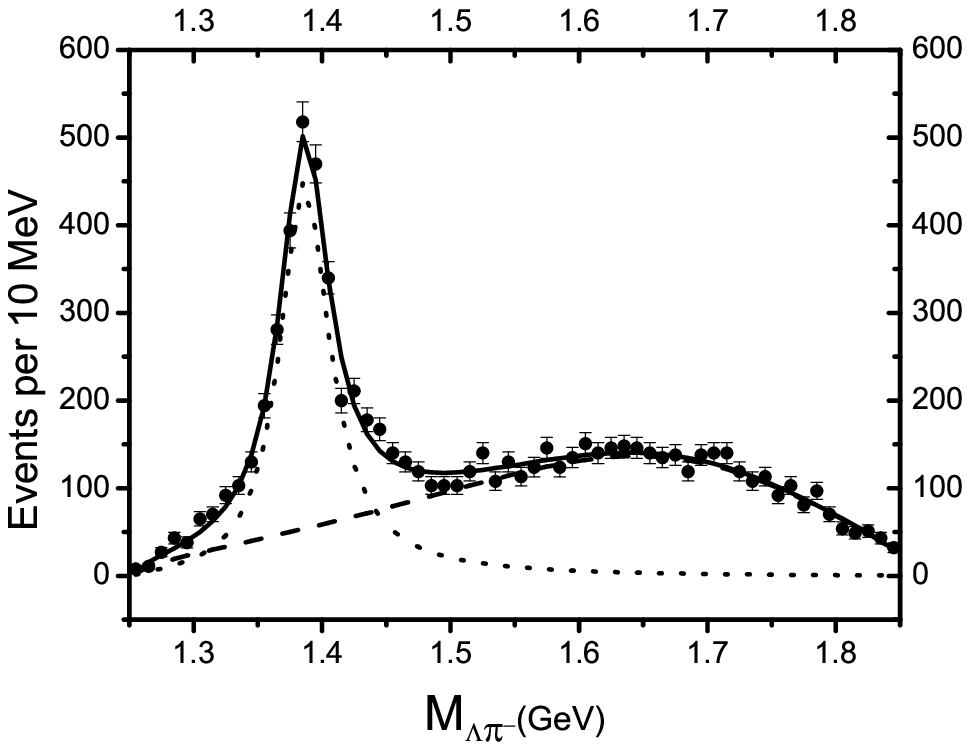}
\includegraphics[width=0.49\columnwidth]{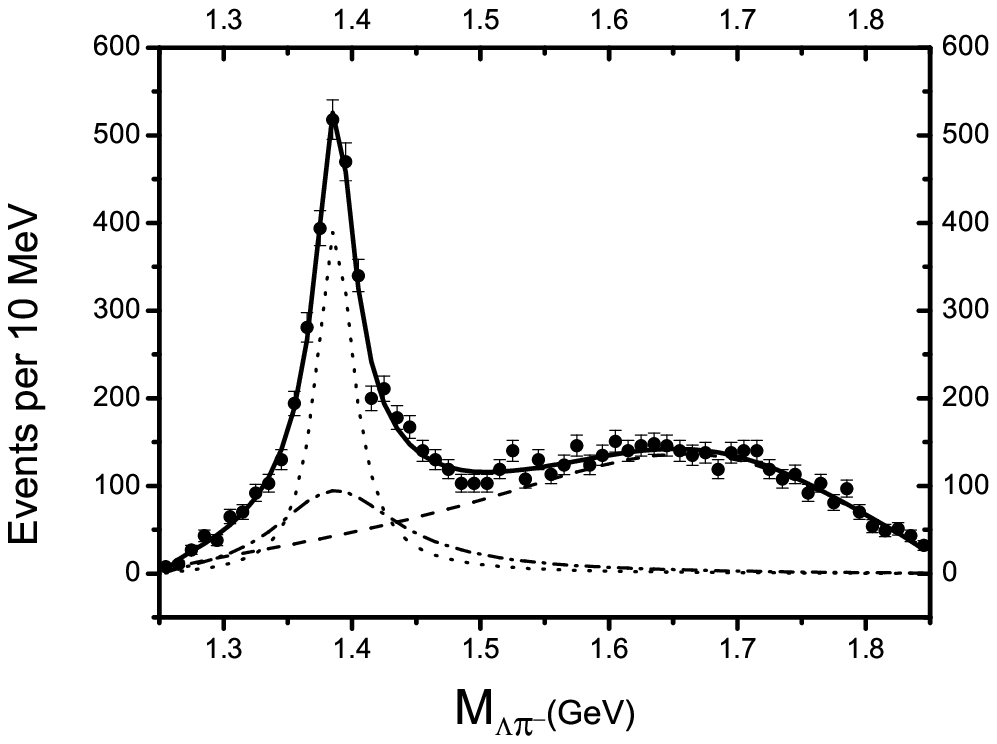}
\vspace{-0.5cm} \caption{Fits to the $\Lambda\pi^-$ mass spectrum
with a single $\Sigma^*$ (left) and two $\Sigma^*$ resonances
(right) around 1385 MeV with fitting parameters listed in Table 1.
The experiment data are from Ref.\cite{pr69} on
$K^-p\to\Lambda\pi^+\pi^-$ with beam momenta around 1.4 GeV.}
\label{f1}
\end{center}
\end{figure}

The results of the fits with a single and two $\Sigma^*$ resonances
around 1385 MeV are shown in Fig.~\ref{f1} and Table~\ref{tab1}
where fitted parameters with statistical errors are given. The fit
with a single $\Sigma^*$ resonance (Fit1) is already not bad. The
fit with two $\Sigma^*$ resonances (Fit2) improves $\chi^2$ by 10.5
compared with the Fit1 for 60 data points with 3 more fitting
parameters. Although this is just a less than $3\sigma$ improvement,
a point favoring Fit2 is that while the single $\Sigma^*$ resonance
in Fit1 has a width larger than the PDG value~\cite{pdg} of $36\pm
5$ MeV for the $\Sigma^*(1385)$ resonance, the narrower $\Sigma^*$
resonance in Fit2 gives a width compatible with the PDG value for
the $\Sigma^*(1385)$ resonance. In the Fit2, there is an additional
broader $\Sigma^*$ resonance with a width about 120 MeV.

\begin{table}[ht]
\begin{tabular}{ c c c c c c c c c  }
\hline \hline    &  $M_{\Sigma^*(3/2)}$ & $\Gamma_{\Sigma^*(3/2)}$\
&  $M_{\Sigma^*(1/2)}$ & $\Gamma_{\Sigma^*(1/2)}$
& $\chi^{2}/ndf$(Fig.1) &$\chi^2/ndf$(Fig.2)\\
\hline
  Fit1     & $1385.3\pm 0.7$       & $46.9\pm 2.5$        &                        &          & 68.5/54  & 10.1/9\\
  Fit2     & $1386.1^{+1.1}_{-0.9}$  & $34.9^{+5.1}_{-4.9}$ & $1381.3^{+4.9}_{-8.3}$ & $118.6^{+55.2}_{-35.1}$
  & 58.0/51 &  3.2/9 \\
\hline \hline
\end{tabular}
\caption{Fitted parameters with statistical errors and $\chi^2$ over
number of degree of freedom (ndf) for the fits with a single (Fit1)
and two $\Sigma^*$ resonances (Fit2) around 1385 MeV. } \label{tab1}
\end{table}

\begin{figure}[htbp]
\begin{minipage}[t]{65mm}
{\includegraphics[width=\columnwidth]{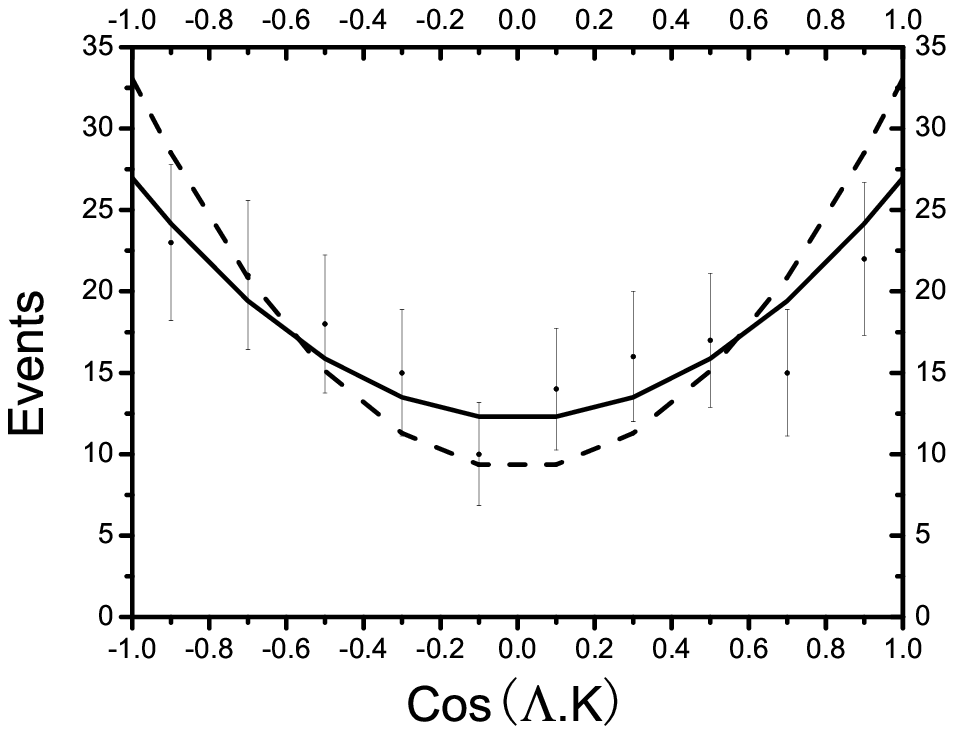}}
\caption{Predictions for the angular distribution for the reaction
$K^-p\to\Lambda\pi^+\pi^-$ at beam momenta around 1.4 GeV by Fit1
(dashed curve) and Fit2 (solid curve), compared with the data from
Ref.~\cite{pr69} .} \label{f2}
\end{minipage}
\hfill
\begin{minipage}[t]{65mm}
{\includegraphics*[width=0.8\columnwidth]{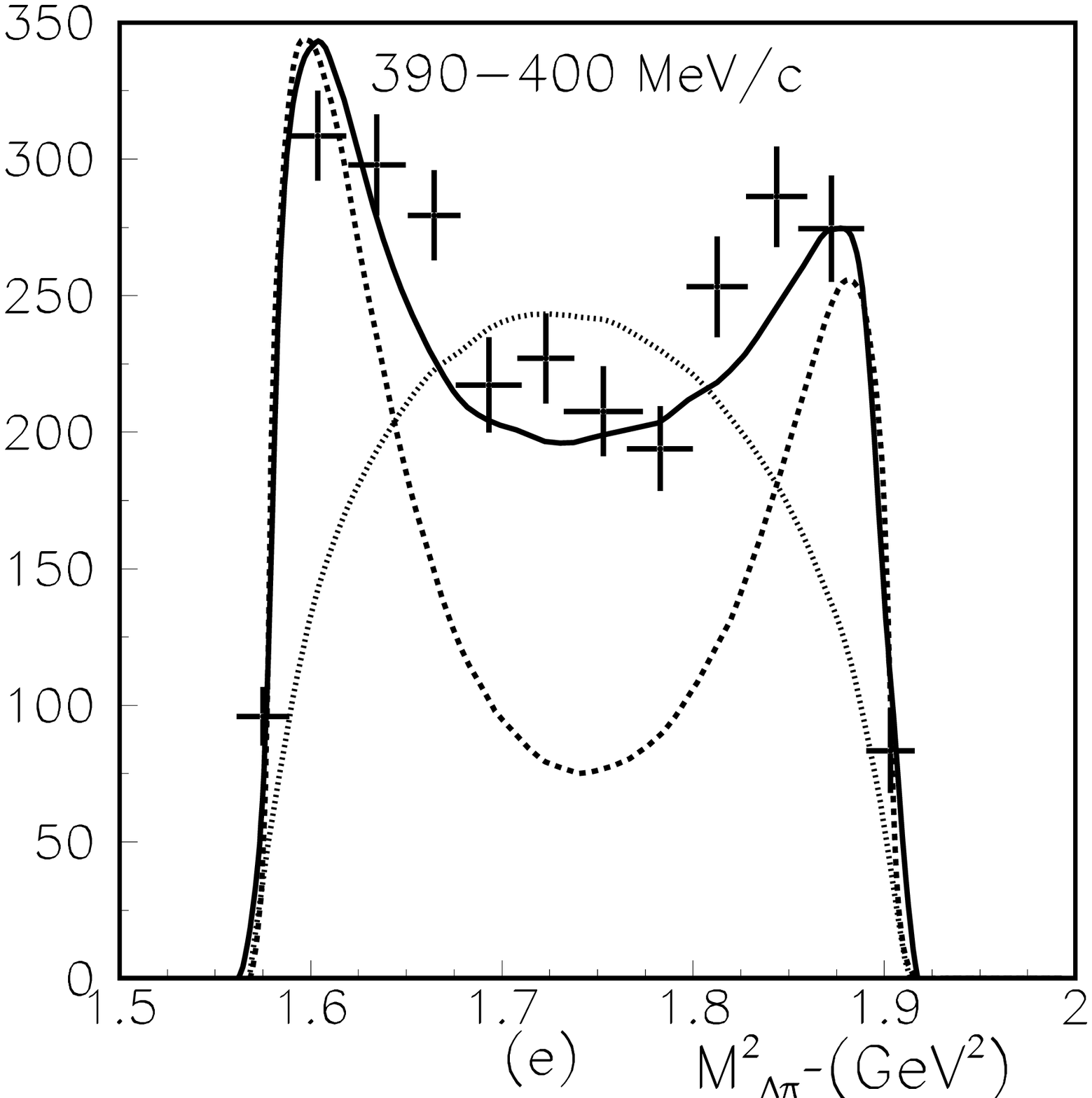}}
\caption{Theoretical $\Lambda\pi^-$ invariant mass  squared
distribution with pure $\Sigma^*(3/2^+)$ (dotted line) and with a
$\Sigma^*(1/2^-)$ in addition (solid line) for the $K^-p\to
\Lambda^{*}(1520)\to \Sigma^*\pi \to\Lambda\pi^+\pi^-$ reaction at
beam momenta around 0.4 GeV, compared with data~\cite{exp}.}
\label{f3}
\end{minipage}
\end{figure}

The preferred assignment of spin $J=3/2$ for the $\Sigma^*(1385)$
resonance in Ref.~\cite{pr69} is demonstrated by the distribution of
the cosine of the angle between the $\Lambda$ direction and the
$K^-$ direction for the reaction $K^-p\to\Lambda\pi^+\pi^-$ with
$M_{\Lambda\pi^-}$ in the range of $1385\pm 45$ MeV and
$cos\theta_{K\Sigma^*} > 0.95$. For a $\Sigma^*$ with $J=3/2$, the
angular distribution is expected to be of the form
$(1+3cos^{2}\theta)/2$; while for a $\Sigma^*$ with $J=1/2$, a flat
constant distribution is predicted. The data~\cite{pr69} as shown in
Fig.2 clearly favor the case of $J=3/2$ if only a single $\Sigma^*$
resonance is assumed. However, the Fit2 with two $\Sigma^*$
resonances with the narrower one of $J=3/2$ and the broader one with
$J=1/2$ reproduces the data even better as shown by the solid curve
in Fig.~\ref{f2}. The results show that the inclusion of an
additional $\Sigma^*(1/2^-)$ besides the well-established
$\Sigma^*(1385)$ $3/2^+$ seems improving the fit to the data of
$K^-$ beam momenta around 1.4 GeV for both $\Lambda\pi^-$ invariant
mass spectrum and the angular distribution although the large error
bars for the angular distribution data make it not very conclusive.

Further stronger evidence is found from a study~\cite{wujj2} on the
data of the $K^-p\to \Lambda^{*}(1520)\to \Sigma^*\pi
\to\Lambda\pi^+\pi^-$ reaction at beam momenta around 0.4
GeV~\cite{exp}. The theoretical $\Lambda\pi^-$ invariant mass
squared distribution with pure $\Sigma^*(3/2^+)$ fails to reproduce
the data as shown by the dotted line in Fig.~\ref{f3}. With a
$\Sigma^*(1/2^-)$ in addition, the data are perfectly reproduced as
shown by the solid line in Fig.~\ref{f3}.

Recently, the LEPS collaboration reported a measurement of the
reaction $\gamma n\to K^+\Sigma^{*-}(1385)$ with linearly polarized
photon beam at resonance region~\cite{hicks}. The beam asymmetry is
sizably negative at $E_\gamma=1.8-2.4 \mathrm{GeV}$, which is in
great contrast to the theoretical prediction of a very small beam
asymmetry~\cite{Oh}. It is found that including an additional
$\Sigma^*(1/2^-$ gives naturally the sizably negative beam
asymmetry~\cite{gaopz}.

\section{5-quark components in other baryons} \label{Sect4}

The 5-quark components in other baryons are also found important. An
admixture of 10-20\% of $qqqq\bar q$ components in the
$\Delta(1232)$ resonance is shown to reduce the well known
under-prediction for the decay width for $\Delta(1232)\to N\gamma$
decay by about half and that of the corresponding helicity
amplitudes about a factor 1.6. The main effect is due to the
quark-antiquark annihilation transitions $qqqq\bar q\to qqq\gamma$,
the consideration of which brings the ratio $A_{3/2}/A_{1/2}$ and
consequently the E2/M1 ratio $R_{EM}$ into agreement with the
empirical value~\cite{riska}. The large contribution of the
quark-antiquark annihilation transitions may also compensate the
under-prediction of the $\pi N$ decay width of the $\Delta(1232)$ by
the valence quark model, once the $\Delta(1232)$ contains $qqqq\bar
q$ components with $\sim$ 10\% probability~\cite{riska1}. The
presence of substantial $qqqq\bar q$ components in the $N^*(1440)$
can bring about a reconciliation of the constituent quark model with
the large empirical decay width of the
$N^*(1440)$~\cite{riska2,JuliaDiaz:2006av}.

From a recent study \cite{xiejj2} of the strong near-threshold
enhancement of $pp \to nK^+\Sigma^+$ cross section, the
extraordinary large coupling of the $\Delta^{*}(1620)$ to $\rho N$
obtained from the $\pi^+ p\to N\pi\pi$ is confirmed. Does the
$\Delta^{*}(1620)$ contain a large $\rho N$ molecular component or
relate to some $\rho N$ dynamical generated state? If so, how about
its SU(3) decuplet partners?  In fact, from PDG
compilation~\cite{pdg} of baryon resonances, there are already some
indications for a vector-meson-baryon SU(3) decuplet. While the
$\Delta^{*}(1620)1/2^-$ is about 85 MeV below the $N\rho$ threshold,
there is a $\Sigma^*(1750)1/2^-$ about 70 MeV below the $NK^*$
threshold and there is a $\Xi^*(1950)?^?$ about 60 MeV below the
$\Lambda K^*$ threshold. If these resonances are indeed the members
of the $1/2^-$ SU(3) decuplet vector-meson-baryon S-wave states, we
would expect also a $\Omega^* 1/2^-$ resonance around 2160 MeV. All
these baryon resonances can be searched for in high statistic data
on relevant channels from vector charmonium decays by upcoming BES3
experiments in near future.

There are many other baryon resonances which are proposed to be
meson-baryon states~\cite{oset}, {\sl i.e.}, $qqqq\bar q$ regrouping
into two colorless clusters.

\section{Conclusion} \label{Sect5}

For the proton, there should be at least about $20\%$ mixture of the
five-quark components in the proton to reproduce its large $\bar
u$-$\bar d$ asymmetry ($\bar d-\bar u\approx 0.12$) and $s$-$\bar s$
asymmetry. To tell the relative importance of meson cloud and
diquark cluster mixtures in the proton, more precise experiments on
the strangeness radius and strangeness magnetic moment of the proton
are needed.

There should be more five-quark components in excited baryons. The
study of $1/2^-$ baryons seems telling us that the $\bar qqqqq$ in
S-state is more favorable than $qqq$ with $L=1$. In other words, for
excited baryons, the excitation energy for a spatial excitation
could be larger than to drag out a $q\bar q$ pair from gluon field.
Whether the $\bar qqqqq$ components are in diquark cluster
configuration or meson-baryon configuration depends on the strength
of relevant diquark or meson-baryon correlations. For $N^*(1535)$
and its $1/2^-$ SU(3) nonet partners, the diquark cluster picture
for the penta-quark configuration gives a natural explanation for
the longstanding mass-reverse problem of $N^*(1535)$, $N^*(1440)$
and $\Lambda^*(1405)$ resonances as well as the unusual decay
pattern of the $N^*(1535)$ resonance. For $\Delta^{*++}(1620)$ and
its $1/2^-$ SU(3) decuplet partners, their SU(3) quantum numbers do
not allow them to be formed from two good scalar diquarks plus a
$\bar q$. Then their $\bar qqqqq$ components would be mainly in
meson-baryon configurations.

Distinctive patterns are predicted by quenched quark models and
unquenched quark models for the lowest SU(3) baryon nonet with spin
parity $J^P=1/2^-$. While the quenched quark models predict the
lowest $1/2^-$ $\Sigma^*$ resonance to be above 1600 MeV, the
unquenched quark models predict it to be around $\Sigma^*(1385)$
energy. By re-examining some old data of the
$K^-p\to\Lambda\pi^+\pi^-$ reaction, it is found that besides the
well established $\Sigma^{*}(1385)$ with $J^P=3/2^+$, there is
indeed some some evidence for the possible existence of a new
$\Sigma^{*}$ resonance with $J^P=1/2^-$ around the same mass but
with broader decay width. There are also indications for such
possibility in the $J/\psi\to\bar\Sigma\Lambda\pi$ and $\gamma n\to
K^+\Sigma^{*-}$ reactions. At present, the evidence is not very
strong. Therefore, high statistics studies on the relevant
reactions, such as $K^-p\to\pi\Sigma^*$, $\gamma N\to K\Sigma^*$,
$\psi\to\bar\Sigma\Sigma^*$ with $\Sigma^*\to\Lambda\pi$ or
$\Sigma\pi$, are urged to be performed by forthcoming experiments at
JPARC, CEBAF, BEPCII, etc., to clarify the situation.

\bigskip
\noindent {\bf Acknowledgements}  I would like to thank D.O.Riska,
B.C.Liu, J.J.Xie, C.S.An, J.J.Wu, F.X.Wei, X.Cao, S.Dulat and
H.C.Chiang for collaborations on relevant issues reported here, and
thank the Institute for Nuclear Theory at the University of
Washington for its hospitality during the completion of this paper.
This work is partly supported by the National Natural Science
Foundation of China (NSFC) under grants Nos. 10875133, 10821063,
10635080, and by the Chinese Academy of Sciences (KJCX3-SYW-N2), and
by the Ministry of Science and Technology of China (2009CB825200).

\end{document}